\documentclass[AMA,STIX1COL]{WileyNJD-v2}

\articletype{Article Type}%

\received{}
\revised{}
\accepted{}
        
\usepackage{comment}
\usepackage{xcolor}
\usepackage{diagbox}
\newcommand{\RNum}[1]{\uppercase\expandafter{\romannumeral #1\relax}}
\newtheorem{Def}{Definition}
\newtheorem{Theo}{Theorem}

\begin{document}

\title{Correlated Data in Differential Privacy: Definition and Analysis}

\author[1]{Tao Zhang}
\author[1]{Tianqing Zhu*}
\author[2]{Renping Liu}
\author[1]{Wanlei Zhou}

\authormark{AUTHOR ONE \textsc{et al}}

\address[1]{\orgdiv{Centre for cyber security and privacy with School of Computer Science}, \orgname{University of Technology
		}, \orgaddress{\state{NSW}, \country{Australia}}}
\address[2]{\orgdiv{School of Electrical and Data Engineering }, \orgname{University of Technology}, \orgaddress{\state{Sydney}, \country{Australia}}}

\corres{*Tianqing Zhu, \email{Tianqing.Zhu@uts.edu.au}}

\presentaddress{15 Broadway, Ultimo, Sydney, New South Wales 2007, Australia}

\abstract[Summary]{Differential privacy is a rigorous mathematical framework for evaluating and protecting data privacy. In most existing studies, there is a vulnerable assumption that records in a dataset are independent  when differential privacy is applied. However, in real-world datasets, records are likely to be correlated, which may lead to unexpected data leakage. 
	In this survey, we investigate the issue of privacy loss due to data correlation under differential privacy models. Roughly, we classify existing literature into three lines: 1) using parameters to describe data correlation in differential privacy, 2) using models to describe data correlation in differential privacy, and 3) describing data correlation based on the framework of Pufferfish. Firstly, a detailed example is given to illustrate the issue of privacy leakage on correlated data in real scenes. Then our main work is to analyze and compare these methods, and evaluate situations that these diverse studies are applied. Finally, we propose some future challenges on correlated differential privacy. }


\maketitle

\section{Introduction}\label{sec-intro}

Over the last decade, the relationship between human and data has never been so inseparable. Meanwhile, the era of big data poses new challenges to human data management, especially in data privacy \cite{Yu2016}.  Privacy preserving data releasing has been adopted by academia and industry to protect individual privacy when datasets are are published to the public \cite{Wang2010,mehmood2016protection}. In order to guarantee data security, data need to be sanitized via privacy mechanisms, such as k-anoymity \cite{sweeney2002k},  l-diversity \cite{machanavajjhala2007diversity}, t-closeness \cite{li2007t}. Among these privacy mechanisms, differential privacy is one of the most promising privacy models to protect data privacy. The notion of differential privacy was firstly proposed by Dwork et al. \cite{dwork2008differential}, which provides a rigorous mathematical framework of defining and protecting privacy. A common method to achieve differential privacy is to add random noise to the output of a query. It ensures that the adversary cannot distinguish the participation of the individual even if the adversary knows the entire background information.

In traditional differential privacy, a weak assumption is that records in a dataset are independent from each other. In practice, however, data in a dataset are usually correlated resulting from the process of data generation. As such, deleting one record will have impacts on other records. Such impacts may reveal more information for the adversary.  Kifer et al. confirms that the correlation between data may disclose more information than expected \cite{kifer2011no}. This finding starts a new topic on how to preserve the privacy of correlated datasets when they are released to the public. Adding  noise to correlated datasets has been proposed as one of methods to guarantee differential privacy. One of the challenges is how to add appropriate noise to preserve data privacy in correlated datasets since adding too much noise to the correlated dataset will degrade data utility, and adding insufficient noise will disclose data privacy. 

Generally, the amount of noise added to correlated datasets depends the extent of data correlation, which is an inherent feature from data generation.  In order to add appropriate noise, data generation or data correlation should be known by the curator or the adversary as background information. Hence, many works modify traditional definition of differential privacy, and add background knowledge of data correlation by correlation parameters and correlation models, to cope with the issue of privacy loss for the correlated dataset.
Pufferfish \cite{Kifer2014} is a flexible privacy model without the assumption that data are independent in a dataset, which can be used to quantify the privacy loss due to data correlation.

In terms of how to describe data correlation, we classify existing research into three streams.
The first stream uses parameters to describe simple data correlation in differential privacy. Chen. et al used the method of multiplying original sensitivity with the number of correlated records, yet it may lead to too much noise \cite{chen2014correlated}. Other correlation parameters are proposed to describe data correlation in differential privacy, including the correlated degree matrix \cite{Zhu2015} and the dependence coefficient \cite{liu2016dependence}. 
The second stream exploits correlation models to describe complex data correlation in differential privacy, such as Gaussian correlation model \cite{yang2015bayesian,chen2017correlated} and Markov chain model \cite{cao2018quantifying,xiao2015protecting}. The last stream is built on the privacy framework, called Pufferfish \cite{Kifer2014}, which is a flexible privacy model to guarantee the data sharing needs and is able to describe simple and complex data correlations. Inspired by Pufferfish, He et al. proposed another privacy model, Blowfish to tune privacy-utility trade-off \cite{he2014blowfish}.



\subsection{Outline and Survey Overview}
Several surveys have been working on differential privacy. The first survey by Dwork \cite{dwork2008differential} summarized notions of differential privacy, mechanisms and some differentially private algorithms for data publishing. Later,  Dwork et al. gave an overview on motivated applications and future directions for data publishing and data analysis \cite{dwork2010differential,Dwork2011}. A book by Dwork presented comprehensive coverage of algorithms maintaining differential privacy against adversaries and differentially private methods for mechanism design and machine learning \cite{Dwork2014}.
Sarwate et al. studied differentially private algorithms for continuous data in signal processing \cite{sarwate2013signal}. Recently, Zhu et al. gave a summary on the data publishing and data analysis underlying differential privacy \cite{Zhu2017}. Damien et al. gave a systematic taxonomy of these variants and extensions of differential privacy
\cite{desfontaines2020sok}. Previous surveys mainly focus on the concepts, theories and development of differential privacy.
Different from existing works, this survey focuses on the issue of privacy leakage on correlated data, which is a vital issue in differential privacy. 
The contributions of this paper are listed as below.
\begin{itemize}
	\item First, we give a summary of existing research on correlated differential privacy, and roughly classify existing research into three research streams: correlation parameters, correlation models and Pufferfish. This helps to understand the characteristics of existing methods on correlated differential privacy.
	\item Second, we compare the advantages and disadvantages, similarities and differences and the application scenarios of methods. This provides a guideline to use correlated differential methods in different scenarios.
	\item Finally, we propose a number of future topics on correlated differential privacy. This gives some sights on new issues and potential methods in correlated differential privacy.
\end{itemize}
 
The rest of this paper is organized as follows. We describe the preliminaries in Section \RNum{2}, and gives examples to illustrate the problem in Section \RNum{3}. Section \RNum{4}   and Section \RNum{5} summarize the studies in differential privacy models by correlation parameters and correlation models, respectively. In Section \RNum{6}, we introduce the framework of Pufferfish. Section \RNum{7} is the future direction, and finally, Section \RNum{8} is the conclusion.

\section{Preliminaries} 
\subsection{Differential privacy} 
Differential privacy is a rigorous privacy model which is widely studied in the last decade. In brief,  $ D $ is a dataset that contains a set of records. Two datasets $D $ and $ D^{'}$ are referred to as neighboring datasets when they differ in one record. 
A query $ f $ is a function that maps records $ r \in \Omega $  to  abstract outputs $ f(D) \in \Omega $, where $\Omega$ is the whole set of outputs. Hence, the dataset is the input and the released information from the mechanism is the output. The relationship can be described as $ f $: $D \xrightarrow{} \Omega $. 

\begin{Def}
	($\epsilon$-Differential privacy) \cite{dwork2008differential} Given neighboring datasets $ D $  and $ D^{'} $, a randomized algorithm $\mathcal{M} $ satisfies \emph{$ \epsilon $-differential privacy} for any possible outcome $ f(D) \in \Omega $,
	\begin{equation}
 Pr[\mathcal{M}(D) \in \Omega] \leq exp(\epsilon) \cdot Pr[\mathcal{M} (D^{'}) \in \Omega  ] 
	\end{equation}	
\end{Def}
where $ \epsilon $ is  privacy budget which determines privacy level. The lower $\epsilon$ represents the higher privacy level. 
\begin{Def}
	(Sensitivity). \cite{dwork2008differential} For a query $f: D \xrightarrow{} \Omega $, and neighboring datasets, the sensitivity of $ f $ is defined as		
\end{Def}
\begin{equation}
 \Delta f= \max \limits_{D,D^{'} } ||f(D)-f(D^{'})||_1
\end{equation}
Sensitivity measures the maximal difference between neighboring datasets. When a dataset is given, sensitivity depends on the type of query $ f $.
\subsection{Differential Mechanisms}
Two common mechanisms are widely used to achieve $\epsilon$-differential privacy: Laplace mechanism \cite{dwork2008differential}  and Exponential mechanism \cite{McSherry2007}. 
\begin{Def}
	(Laplace mechanism). Given a query $ f $:$ D \xrightarrow{}\Omega $ over the dataset $ D $, Laplace mechanism satisfies $\epsilon$-differential privacy if,
	\begin{equation}
	\mathcal{M}(D)=f(D)+Laplace(\Delta / \epsilon) 
	\end{equation}
\end{Def}

where $ Laplace (\cdot) $ denotes Laplace noise which is drawn from a Laplace distribution with the probability density function $ p(x| \lambda)=\frac{1}{2 \lambda}e^{-|x|/ \lambda} $, where $ \lambda $ depends on privacy budget and sensitivity.
\begin{Def}
	(Exponential mechanism). Given a score function $ S(D,\phi) $ of a dataset $ D $, exponential mechanism $\mathcal{M}  $ satisfies $\epsilon$-differential privacy if
	\begin{equation}
	 \mathcal{M} (D)= \left(return \ \phi \propto  exp(\frac{\epsilon S(D,\phi) } {2 \Delta f}) \right) 
	\end{equation}
\end{Def}
where the score function $ S(D,\phi) $ is used to evaluate the quality of an output $ \phi $ for a query $f$.
 Exponential mechanism describes that the probability of returning $ \phi $ increases exponentially with the increase in the value of $ S(D,\phi) $. 
\subsection{An analysis of data correlation}
In this section, we will introduce some most frequent studied correlations in the literature. Many types of correlations are in real-world datasets, and the correlation is assumed to be known by the curator and the strong adversary. Generally, data correlation can exist in one dataset or in multiple datasets. In the first case, data correlation can disclose more information when the dataset is published from one entity. In the second case, personal information may appear in different entities. For instance, people would like to share their information in different social applications (e.g., Twitter or Facebook), and these information can be shared to the third party via social applications at the same time. 

\subsubsection{Direct correlation}
Direct correlation occurs when the curator has access to all knowledge of data correlation. For example, $A, B, C$ are records in a dataset, and direct correlation between these records can be expressed as $ A+B=C $ or $ A*B=C $, etc. Direct correlation is deterministic; hence, it is relatively easy for the curator to handle. Direct correlation is simple data correlation.

\subsubsection{Indirect correlation}
Different from the direct correlation, indirect correlation is more complex. It is non-deterministic and thus can not defined the correlation as a formula. Indirect correlation is complex data correlation.
\paragraph{Temporal correlation}
A dataset with temporal correlation is generated by the predefinition of a time interval, and continuous released records falling into this time interval are regarded as correlated by time. Continuous generated data in the real world tend to be temporally correlated, like the dataset of user locations described in the example above. One characteristic of temporal correlation is that all records are usually correlated, which means the first record in the dataset may have an impact on the last record. The extreme case described in Section \RNum{3} is an example of such a case. And because all records in the temporally-correlated dataset are related, some studies have solely focused on differential privacy given temporal correlation \cite{cao2018quantifying,xiao2015protecting,bozkir2020differential}.
\paragraph{Attribute correlation}
Attribute correlation refers to correlations that can be revealed through a particular attribute, i.e., when the value of two or more records is the same or similar. In reality, there are many attributes that can create correlations in real-world datasets, and many real-world datasets contain those attributes. For example, addresses, which are common to social network and ancestry datasets, are an attribute that can be used to identify members of the same family.


\section{Problem statement}

In this section, we show the issue of privacy loss due to data correlation.
Most previous works assume that all records in the dataset are independent. Based on this assumption, differential privacy claims that it can limit the probabilistic inference when the attacker knows the whole information but one record. However, records in a real-world dataset are often correlated with each other, and it is likely to breach the privacy in a dataset when the adversary knows all but one record and the knowledge of data correlation. Here, we give an example to illustrate the temporal correlation in a dataset, and how data correlation will degrade privacy level.
\section{An example to illustrate data correlation}


Considering the scenario of a traffic monitoring application, data of user mobility are collected by a trusted server continuously. In this scenario, one typical correlated dataset is generated - temporal correlated dataset. This type of dataset is generated continuously in a time interval, and released records are correlated due to time correlation. 
Users in a monitoring area are likely to have social relationships - perhaps friends or couples. Due to social relationships, location information is the same for some users during a period of time, and thus users' location information have some correlation in a dataset.

\begin{table}[!htbp]
	\centering 
	\caption {Users' locations at different time points}
	
	\begin{tabular}{|c|c|c|c|c|}
		\hline
		\diagbox{user}{t}&{1}&{2}&{3}&{4}\\ 
		\hline 
		$ u_{1} $&$ loc_{2} $&$ loc_{2} $&$ loc_{3} $&$ loc_{4} $\\
		\hline
		$ u_{2} $&$ loc_{2} $&$ loc_{2} $&$ loc_{3} $&$ loc_{4} $\\
		\hline
		$ u_{3} $&$ loc_{1} $&$ loc_{4} $&$ loc_{5} $&$ loc_{2} $\\
		\hline
		$ u_{4} $&$ loc_{4} $&$ loc_{5} $&$ loc_{2} $&$ loc_{5} $\\
		\hline 
	\end{tabular}
\end{table}

\begin{table}[!htbp]
	\centering 
	\caption {Sum counts of users' locations}
	
	\begin{tabular}{|c|c|c|c|c|}
		\hline
		\diagbox{loc}{t}&{1}&{2}&{3}&{4}\\ 
		\hline 
		$ loc_{1} $&1&0&0&0\\
		\hline
		$ loc_{2} $&2&2&1&1\\
		\hline
		$ loc_{3} $&0&0&2&0\\
		\hline
		$ loc_{4} $&1&1&0&2\\
		\hline
		$ loc_{5} $&0&1&1&1\\
		\hline 
	\end{tabular}
\end{table}

 As shown in the Table \RNum{1}, users' location are given in different time points. 
 In Table 1, we can note that $ user_{1} $  and $ user_{2} $ have the same location from the time point $ t=1 $ to $ t=4 $. The reason could be $ user_{1} $  and $ user_{2} $  are family members, they are likely to have the same track in a time period.  In this case, changing the location of $ user_{1} $ will also change the location of $ user_{2} $, hence the records of $ user_{1} $ and $ user_{2} $ are referred to as correlated records.

 Table \RNum{2} shows the sum of true counts with regard to user locations.
 When Laplace mechanism is applied in this case, the
 amount of Lap(1/$\epsilon $) noise is added to perturb each count in Table \RNum{2} so that released information can achieve  $ \epsilon $-DP at each time point. However, if the attacker knows the relationship between $ user_{1} $  and $ user_{2} $, the attacker can infer the location of $ user_{1} $ and  $ user_{2} $. 
As a result, when the count of users location is released, the privacy of users location is unlikely to  satisfy $\epsilon$-differential privacy because the same count is considered to be released two times.  Lap(2/$\epsilon $) noise should be added to the query result in order to hold differential privacy in the dataset.

Based on the example above, we can find that data correlation can exist in a dataset. The change of one record can have an impact on other records, and it also leads to changes on the query response. This example proves that correlated data in a dataset will leak more information to the adversary when using differential privacy, and hence degrade the privacy level. Intuitively, one method is to inject more noise to the correlated dataset. The amount of noise added to the correlated dataset depends on the degree of correlated information in a dataset. This situation reveals that the level of challenge faced in dealing with the trade-off between data utility and data privacy. 


\section{Correlation parameters in differential privacy}

In this section, we will introduce some works that use correlation parameters to describe data correlation on correlated datasets in differential privacy.
After data correlation is measured by correlation parameters, an appropriate amount of noise can be quantified to add in differential privacy, and thus privacy can be guaranteed with a desirable trade-off between data privacy and data utility.



\subsection{The number of correlated records} 
One of the simple method to describe data correlation was proposed by Chen et al. \cite{Chen2014}.  The correlation parameter $ k $ is used to measure the extent of correlated data. A dataset $ D $ with a correlation parameter $ k $ means that the maximum number of correlated records in the dataset is $ k $.  The correlation parameter $ k $ is assumed to be known by the curator or the strong adversary. After multiplying the original sensitivity with the number of $ k $ correlated records,  any $ \frac{\epsilon}{k} $ differentially private mechanism also satisfies $ \epsilon $ differential privacy when the number of $ k $ correlated records are in the dataset. 

$ \textit{Analysis:}$ The advantage of this method is the simplicity and easy to implement. However, when a batch of correlated records are in a dataset, a large amount of noise will be added in the output since the correlated parameter $ k $ cannot describe data correlation accurately. Hence, this will lead to a severe degradation in dataset utility. 

\subsection{Dependence coefficient}
Liu et al. studied data correlation with a new definition of dependent differential privacy, which considers more background knowledge of data correlation described by a correlation parameter $ \rho_{ij}$, called the dependence coefficient \cite{liu2016dependence}. Firstly, the definition of dependent differential privacy (DDP) is given below.

\begin{Def}
	($\epsilon$-Dependent Differential Privacy) A mechanism $  \mathcal{M}$ gives $\epsilon$-dependent differential privacy for any pairs of dependent neighboring datasets $ D(L,R) $ and $ D'(L,R) $ and any possible outcomes $ \Omega $, if the mechanism $\mathcal{M}$ satisfies
	\begin{equation}
	 \max \limits_{D(L.R),D'(L,R)} \frac{P(\mathcal{M}(D(L,R)=\Omega))}{P(\mathcal{M}(D'(L,R)=\Omega))} \leq exp(\epsilon)
	 \end{equation}	
\end{Def}

\noindent where $L$ is the number of correlated records and $ R $ is the probabilistic dependence relationship between the records.

In the definition of $\epsilon$-dependent differential privacy, we note two differences from the traditional differential privacy. One is the probabilistic dependence relationship is specified in the dataset and the other is the size of correlated records is specified in the dataset. From the definition of DDP, we see that the DDP can guarantee the data privacy and defend against the attacker who even has the background information of probabilistic dependence $ {R} $ between records. More specific, dependent sensitivity includes two parts: the sensitivity caused by the modification of the record itself $ \triangle D_{j} $ and the sensitivity induced in other records $ \rho_{ij} \triangle D_{ij} $. The dependence coefficient $ \rho_{ij} \in [0,1] $ serves as a metric to evaluate the extent of the dependent relationship between tuples. 

\begin{Def}
	(Dependent sensitivity) For a query $ f $, dependent sensitivity is defined over a dependent dataset $ D $ as,
	\begin{equation}
	 DS^{f} = \max \limits_{i} \sum_{j=C_{i1}}^{C_{iL}    }\rho_{ij} \Delta f_{j}  
	\end{equation}
	where $C_{i1}$,...,$C_{iL}$ denotes $ L $ records that are dependent with $ i $-th record and $ \rho_{ii} = 1 $. 
	$ DS^{f} $ denotes dependent sensitivity of a query $ f $ over all records in the dataset $  D  $ caused by the modification of one individual record $ D_{i} $. 
\end{Def}

$ \textit{Analysis:}$ The advantage of dependence coefficient or DDP is that this correlation parameter is able to measure the degree of data correlation. While 
the effectiveness of dependence coefficient depends on how well the correlation between records can be described and computed. For example, when the correlation in a dataset is exactly known by the curator, this method can model it well. When data correlation is unknown or partial known, the accuracy of dependence coefficient may be overestimated or underestimated.
\subsection{Zhao-Dependent Differential Privacy}
Another kind of dependent differential privacy studied in \cite{zhao2017dependent}, which we refer to as Zhao-DDP in this paper. The goal of Zhao-DDP is to prevent the adversary from inferring the user's information with the combination of correlated records and query responses. The definition of Zhao-DDP is given below.
\begin{Def}
	(Zhao-$ \epsilon $-Dependent Differential Privacy) A mechanism $ \mathcal{M} $ provides $ \epsilon $-DDP, if for any neighbouring datasets and any possible outputs $ \Omega $, we have
	\begin{equation}
	 \frac{\mathbb{P}[\mathcal{M}(x_{i},x_{K},X_{\overline{K}}) \in \Omega]} {\mathbb{P}[\mathcal{M}(x'_{i},x_{K},X_{\overline{K}})  \in \Omega]} \leq e^{\epsilon_{c}},{\forall}i,K,x_{i},x'{i},x_{K}, \Omega 
	\end{equation}
	where $ i \in \{1,...,n\}, K \subseteq  [\overline{ i }]=\{1,...,n\} \backslash \{i\}, \overline{K}=[\overline{i} ]\backslash K$; $ \epsilon'$ is a segmented linear function of traditional $ \epsilon $-differential privacy.
\end{Def}
When calculating the conditional probabilities, the correlation knowledge is needed from the curator. Comparing with differential privacy, we can find that K iterates through all subsets of $ [\overline{i}]={1,...,n} \backslash \{i\} $ to bound $ \frac{\mathbb{P}[\mathcal{M}(x_{i},x_{K},X_{\overline{K}}) \in \Omega]} {\mathbb{P}[\mathcal{M}(x'_{i},x_{K},X_{\overline{K}})  \in \Omega]} $, while $ \epsilon $-DP bounds $ \frac{\mathbb{P} [\mathcal{M}(x_{i},x_{[\overline{ i }]}) \in \Omega ] }{\mathbb{P} [\mathcal{M}(x'_{i},x_{[\overline{ i }]}) \in \Omega ]} $. 

$ \textit{Analysis:}$ Comparing with the DDP in \cite{liu2016dependence}, Zhao-DDP considers more correlation information $ \mathbb{P}[X_{\{1,...,n\} \backslash \{X_{K}\} } |X_{K}  ]$, while DDP considers the correlation information $ \mathbb{P}[X_{\{1,...,n\} \backslash \{i\} } |X_{i}  ]$.

\subsection{Correlated degree matrix}
Zhu et al. used the correlation parameter, correlated degree matrix to describe data correlation  \cite{Zhu2015}. 
In real-world datasets, the extent of correlation between records is different. For example, some records are fully correlated, which means that these records are same records. Some records are partially correlated, which means that changing one record has a probability to change other related records. When the generation of data is not known by the curator or the data correlation is not easy to specify, it is efficient to denote the relation between records with the method of Pearson correlation. With this method, the extent of the impact of a record on another record can be quantified and it is defined as the correlated degree in \cite{Zhu2015}.
With the notion of correlated degree, correlated sensitivity is proposed and defined in the correlated dataset. The definition of correlated sensitivity is given below.
\begin{Def}
	(Correlated sensitivity) Correlated sensitivity for a query $ f $ is defined as, 
	\begin{equation}
	 CS_{q}=\max \limits_{i \in q} \sum_{j=0}^{n} |\delta_{ij}| \{	 \lVert (f(D^{j})-f(D^{-j}) \rVert_{1}    
	\end{equation}
	where $ D_{j} $ and $ D_{-j} $ are neighboring datasets that differ in record $ j $; $ q $ is a set of records; $ \theta_{ij} $ is correlated degree between record $ i $ and record $ j $.  Correlated sensitivity describes the maximal impact on all records in the dataset due to the deletion of one record.	Then, the correlation between records can be expressed with the correlated degree and formed into a correlated degree matrix to show all relationships between records. 
\end{Def}


$ \textit{Analysis:}$ The advantage is that this method can be applied in many cases when there is no special data correlation known by the curator. This is because Pearson correlation can indicate the extent to which records are linear correlated without any prior knowledge of data generation. However, Pearson correlation is a method to evaluate linear relationship between records, hence it may not model the correlation accurately in some cases.

\begin{table}[!htbp]
	\centering 
	\caption { Comparison of correlation parameters underlying differential privacy}
	\begin{tabular}{ m{1.5cm} | m{4cm} | m{4cm} | m{5cm}  } 
		\hline

		Correlation parameter  & Sensitivity& Advantage& Challenge\\
		\hline
		$ k $ \cite{Chen2014}& $k \epsilon$& It is easy to compute.& It may introduce a large amount of noise to the output. \\
		\hline
		$ \rho_{ij} $ \cite{liu2016dependence}&$DS^{f} = \max \limits_{i} \sum_{j=C_{i1}}^{C_{iL}    }\rho_{ij} \Delta f_j  $&  The correlation between record $ i $  and record $ j $ can be  presented clearly. & The utility of correlated dataset  depends on how well the dependence coefficient is computed.  \\
		\hline
		$ \epsilon' $\cite{zhao2017dependent} & $ \Delta f $ & It considers all possible cases  of data correlations.& The data correlation is not presented clearly with the correlation parameter.\\
		\hline
		$ \delta_{ij} $\cite{Zhu2015}&$CS_{q}= \max \limits_{i \in q} \sum_{j=0}^{n} |\delta_{ij}| \{	 \lVert (f(D^{j})-f(D^{-j}) \rVert_{1}    \} $& Correlated degree is able to measure the degree of data correlation.& Calculating correlated matrix degree  is computational comparing with other methods.\\
		\hline 
	\end{tabular}
\end{table}

\subsection{Discussion of correlation parameters}
Above methods show how to describe data correlation with correlation parameters in different settings for correlated datasets, and we make a comparison of these method in Table \RNum{3}. We can note that most these methods need more background knowledge of data generation. The background knowledge is the number of correlated records, and it is not enough to calculate the exact correlations, leading to a higher noise level \cite{Chen2014}. In \cite{liu2016dependence}, the background knowledge is the number of correlated records $ L $  and the probabilistic dependence relationship $ {R} $ between the records. However, the effectiveness of this method relies on how well the correlation between records can be modeled and computed by the probabilistic dependence relationship. It is not easy to compute dependent coefficient accurately unless the probabilistic models of the data is known. 

When the curator has no knowledge of how the data generated, the method proposed in \cite{Zhu2015} can help identify data correlation. Comparing with the method in \cite{liu2016dependence}, the method in \cite{Zhu2015} may not have a better performance. This is because in \cite{Zhu2015}, the sensitivity measures the effect on all records in the dataset according to the Pearson correlation, which may not describe data correlation accurately as the method in \cite{liu2016dependence}.
The above analysis shows that the background knowledge of how data are generated or correlated is essential when addressing the issue of privacy leakage on correlated data. 
Usually, with more background information of data correlation, such as \cite{liu2016dependence,zhao2017dependent}, the correlation can be computed more precisely, leading to a better performance in terms of noise level or data utility. In summary, the effectiveness of each method depends on the background knowledge known by the curator for correlated datasets.

\section{Correlation models in differential privacy}  
In this section, we introduce two widely used models to describe complex data correlations: Gaussian correlation model and Markov chain model. 
In the previous section, we introduce some correlation parameters to describe data correlation for simple correlated datasets. However, it may still be difficult to measure some complex data correlations, like social network datasets and temporal correlated datasets. In this paper, simply correlation refers to the correlation that can be described by correlation parameters, and complex correlation refers to the correlation that is difficult to be measured by correlation parameters and measured by correlation model.

\subsection{Gaussian correlation model}
Gaussian correlation model is proposed to describe the complex data correlation and quantity the privacy loss in a new privacy model, called Bayesian differential privacy (BDP) \cite{yang2015bayesian}. First, we give the definition of Gaussian correlation model as,
\begin{Def}
	(Gaussian correlation model) Let $G(x,W)$ be a weighted undireted graph, where the vertex $ x_{i} \in X $ denotes the record $ i $ in $ X $ and the weight $ w_{ij} $  denotes the correlation between records $ i $ and $ j $. Let $ \mathbf{W} = (w_{ij}) $ be weighted adjacent matrix which contains all weights; $\mathbf{D}=diag(w_{1},...w_{n}) $ be the diagonal matrix of $ G(x,W) $ where $ w_{i}=\sum_{j \neq i}w_{ij} $; $\mathbf{L}=\mathbf{D}-\mathbf{W}$ be the Laplacian matrix of $ G(x,W) $, 
\end{Def}

\begin{equation}
\mathbf{L}=\mathbf{D}-\mathbf{W}=\left(\begin{array}{cccc}w_{1} & -w_{12} & \cdots & -w_{1, n} \\ -w_{12} & w_{2} & \cdots & -w_{2, n} \\ \vdots & \vdots & \ddots & \vdots \\ -w_{1, n} & -w_{2, n} & \cdots & w_{n}\end{array}\right)
\end{equation}
The pair $ (x,\mathbf{L}) $ is called Gaussian correlation model, denoted as $ G(x,\mathbf{L}) $. The conditional joint probability of $\mathbf{x}_{-i}=\mathbf{x}_{[n] \backslash\{i\}}$ denoted as,
\begin{equation}
p\left(\mathbf{x}_{-i} | x_{i}\right) \propto \exp \left(-\frac{\mathbf{x}^{\mathrm{T}} \mathbf{L} \mathbf{x}}{2}\right)
\end{equation}
With Gaussian correlation model, unknown correlation between records can be described, and maximum correlated data can be computed. Gaussian correlation model is often used with Bayesian differential privacy. Overall, the main idea of Bayesian differential privacy is to connect the uncertain query answer with given records in a Bayesian way. The definition of Bayesian differential privacy is given below,

\begin{Def}
	(Bayesian Differential Privacy) \cite{yang2015bayesian}. Given an adversary $ \mathcal{A}=\mathcal{A}(i,\mathcal{K}) $ and a randomized perturbation mechanism $\mathcal{M} (x)=Pr(r\in \Omega|x)  $ on the dataset X, Bayesian differential privacy leakage of $ \mathcal{M} $ related to $ \mathcal{A} $ is 
	\begin{equation}
	 BDPL_{\mathcal{A}}(\mathcal{M}) = \sup \limits_{a,b,X_{\mathcal{K}},\Omega}  log\frac{p(\mathcal{M}(X) \in \Omega|X_{i}=a,X_{\mathcal{K}})}{p(\mathcal{M}(X) \in S|X_{i}=b,X_{\mathcal{K}})}  
	\end{equation}
	where $ \mathcal{K} \subset [n] \backslash \{i\}$ be a tuple set and $ \mathcal{A}(i,\mathcal{K}) $ denotes the adversary with knowledge $ \mathcal{K} $ to attack $ x_{i} $. Then we say $ \mathcal{M} $ satisfies $\epsilon$-Bayesian differential privacy, if 
	\begin{equation}
	 \sup \limits_{ \mathcal{A}} BDPL_{\mathcal{A}} \leq \epsilon 
	\end{equation}	
\end{Def}
Bayesian differential privacy leakage shows the largest difference between $ Pr(r \in S|x_i,x_K) $ and $ Pr(r \in S|x_i^{'},x_K) $ and the leakage is bounded by $ \epsilon $. In the definition of Bayesian differential privacy, the background knowledge is  $ x_{\mathcal{K}} $, rather than $ x_{-i} $ in the differential privacy, which means the adversary in Bayesian differential privacy is weaker than the adversary in differential privacy. However, a weaker adversary may have a greater risk in the Bayesian differential privacy, which depends on the prior and posterior of the data distribution.

$\mathbf{BDP\ vs\ DP }$ There are two cases when Bayesian differential privacy is equivalent to differential privacy: 1) The data are independent in the dataset and the adversary has full knowledge of the dataset except the object of its attack; and 2) The adversary has full knowledge of the dataset except the object of its attack and the correlation between records.

$ \textit{Analysis:}$ 
To describe data correlation in the setting of Bayesian differential privacy, Gaussian correlation model is used to measure data correlation. Advantages of Gaussian correlation model include: 1) Any arbitrary correlation between records can be described by a weighted network with an arbitrary topology structure; 2) Gaussian correlation model assumes that the joint distribution of all records are Gaussian distribution. Since Gaussian distribution is easy to compute, the conditional distribution of some records when given other records can be easy to obtain; and 3) Gaussian correlation model can describe both infinite continuous data and discrete data. Due to these pros,  Gaussian correlation model is suitable to describe data correlation in Bayesian differential privacy, which fits the background knowledge that the adversary partially knows  knowledge of individuals, and the unknown individuals can be estimated by the Bayesian theorem.

\paragraph{Applications of Gaussian correlation model in BDP}


Some works studied correlated data in real-world applications based on Bayesian differential privacy. For example, Gaussian correlation model is used to describe the correlated data under the method of Bayesian differential privacy in Mobile CrowdSensing (MCS) \cite{chen2017correlated}. 
Mobile CrowdSensing is a new sensing paradigm that people can use personal mobile devices to collect data from the surrounding environment \cite{ganti2011mobile}.
Data collected from some mobile applications, such as traffic monitoring and advertisement delivering are from personal devices, and these information are likely to be correlated and leads to information leakage. In \cite{chen2017correlated}, the correlated records come from the correlated group that is divided according to the relationship among participants, and the probabilistic relationship among sensing data records is modeled by the Gaussian correlation model. Furthermore,  Gaussian correlation model is used to describe the correlation structure among sensing data with different prior knowledge. Besides, Liu et al. analyzed the issue of location privacy preserving caused by the effects of temporal and spatial correlations based on Bayesian Geo-indistinguishability \cite{liu2019protecting}. 
Another application uses Gaussian correlation model to describe data correlation in the game theory \cite{wu2017game}.  Data correlation can exist in multiple datasets. To preserve privacy in multiple datasets, Wu et al. constructed a game model of multiple players (or publishers) to preserve the data privacy by controlling the privacy parameters of publishers. In multiple correlated datasets, the privacy of a dataset not only depends on its privacy budget, but also depends on the privacy budget of other datasets. Gaussian correlation model is used to describe the background knowledge of the correlation between multiple datasets.

\subsection{Markov chain model}
Markov chain model is a stochastic model used to describe a sequent of possible events. One feature of this model is that the probability of each event depends only on the state attained in the previous event. Due to this, Markov chain is widely used in modeling user mobility \cite{gambs2012next,mathew2012predicting}.

A Markov Chain contains two components: states and transitions. More precisely, $ P = {p_{1},...,p_{n}} $ denote a set of states, in which each state corresponds to a value and a current value only depends on the previous one. 
A set of transitions, such as $ t_{ij} $ denotes the probability of moving from state $p_i$ to state $ p_j $. If an individual move from a state to an occasional position before returning to this state, then the transition from a state to itself may occur. The sum of the probabilities in each row of the transition matrix is 1. Here, we give an example of location data that is modeled by Markov chain.

\begin{table}[!htbp]
	\centering 
	\caption {Transition matrix}
	\begin{tabular}{m{1cm}|m{1cm}|m{1cm}|m{1cm}}

		         &$loc_{1}$ & $loc_{1}$& $loc_{1}$\\
		\hline
		$loc_{1}$& 0.1& 0.1& 0.8 \\
		\hline
		$loc_{1}$&0.2& 0.3 & 0.5 \\
		\hline
		$loc_{1}$&0.6& 0.6 & 0.2
	\end{tabular}
\end{table}
In Table \RNum{4}, the first column denotes the time point $ t $, and the first row denotes the time point $ t+1 $. We can note that $ Pr(l^{t}|l^{t+1}) =0.6$, which means that one user is at $ loc_3 $ at time point $ t $, then the probability of being $ loc_1 $ at time point $ t+1 $ is 0.6.

\paragraph{Application of Markov chain model}
One popular application of Markov chain model is on temporal correlated datasets.
When users' locations  are continuously recorded,  these records can be considered as a temporal correlated dataset. 
Cao et al. studied the potential privacy loss under the temporal correlated dataset with a traditional mechanism \cite{Cao2018}. The background knowledge includes the individual information except the attacker's object and the temporal correlation.  The parameters of Markov chain can be formed into a transition matrix to describe temporal correlation. Due to temporal correlation, temporal privacy leakage comes, and it is defined as temporal privacy leakage.
\begin{Def}
	(Temporal privacy leakage) Let $ D_{\mathcal{K}}^t $ be the tuple knowledge of the adversary $ A_i $. Temporal privacy leakage of $ \mathcal{M}^t $ for the $ A_i $ is defined as follows.
	
	\begin{equation}
 TPL(A_i,\mathcal{M}^{t})=\sup \limits_{l_i^t,l_t^{t'},r^1,...r^T}log\frac{Pr(r^1,...,r^T|l_i^t,D_{\mathcal{K}}^t)}{Pr(r^1,...,r^T|l_i^{t'},D_{\mathcal{K}}^t)} 
	\end{equation}
	where $ D_{t} $ and $ D_{t}^{'} $ are neighboring dataset. $ l_i^t $ and $ l_i^{t'} $ are two different values of user $ i $'s data at time $ t $ and we have $ D^{t}=D_{\mathcal{K}}^t \cup \{ l_i^t\} $ and $ D^{t'}=D_{\mathcal{K}}^t \cup \{ l_i^{t'}\} $. 
	Temporal privacy leakage includes backward privacy leakage and forward privacy leakage. Dividing the temporal privacy leakage in the $ r^t $, then we can have backward privacy leakage and forward privacy leakage. The analysis shows that backward privacy leakage is likely to accumulate from previous privacy leakage and forward privacy leakage increases with future release. 
\end{Def}

Another work about temporal correlation is related to the location privacy. The background knowledge of data correlation is also modeled through the Markov chain and the neighboring dataset and sensitivity are redefined to fit the applicable problem \cite{xiao2015protecting}.
Let $ p_t^- $ be the prior probability of a user's location at time $ t $.  $ \delta $-location set is a set of minimum number of possible locations  that the sum of prior probabilities is no more than $ 1-\delta $, and the equation is given below.

\begin{equation}
\Delta X_t=min \{s_i| \sum_{s_i} p_t^- [i] \leq 1-\delta\}  
\end{equation}
The goal of  $ \delta $-location set is used to form a dataset that reflects a set of probable locations the user might appear, which is equivalent to the dataset of outputs in differential privacy. 
We can note that the difference from traditional differential privacy lies in the neighboring dataset. The neighboring dataset in the new definition is any possible location $ x_1 $ and $ x_2 $ in the  $ \delta $-location set. This definition states the output of location $ z_t $ is differentially private at time $ t $ for continual released locations under temporal correlations. Moreover, due to the modification of neighboring dataset, two dimensional space turns into multidimensional space. Based on the notion of convex hull, the sensitivity hull is proposed to capture the geometric meaning of sensitivity.

\begin{table}[!htbp]
	\centering 
	\caption { Comparison of correlation models underlying differential privacy}
	
	\begin{tabular}{m{2cm}|m{2cm}|m{2cm}|m{4cm}}
		\hline
		
		Correlation model & Direction& Circulation& Applied situation\\
		\hline
		Gaussian correlation model& Directed graph& Acyclic& It can describe induced correlation.  \\
		\hline
		Markov correlation model&Undirected graph& Cyclic & There is no direction for data correlation and can represent cyclic dependencies.  \\
		\hline
		
	\end{tabular}
\end{table}

\subsection{Discussion of correlation models}

The advantage of correlation model is that it can model complex data correlation and can be used to express in the form of the posterior probabilities with other background knowledge expressed in the form of prior probabilities. The comparison of correlation models is in the Table \RNum{5}. The experiments \cite{chen2017correlated} shows the proposed perturbation mechanism based on Bayeisan differential privacy introduces less noise to the query results of correlated sensing data, comparing with Zhu's scheme \cite{Zhu2015} and Chen's scheme  \cite{chen2014correlated}. 
For simple data correlation, adopting correlation parameters to model data correlation is easy to compute and to protect data privacy with a high utility of data. For example, the background knowledge of correlation is described by dependent coefficient \cite{liu2016dependence} and by correlated degree matrix \cite{Zhu2015}. For complex data correlation, Gaussian correlation model and Markov correlation model are considered as powerful methods to describe complex correlated dataset. For temporal correlated datasets, Markov chain model is a suitable method to present data correlation and guarantee the data privacy. Due to the advantage of correlation model, more and more works tend to adopt it to describe the correlated dataset. However, the best choice of correlation model depends on the specific type of correlations. 


\section{The framework of Pufferfish}

In this section, we first introduce Pufferfish and its variant, and then give mechanisms for Pufferfish.
Pufferfish is a privacy framework that is proposed to cope with the issue of privacy leakage in correlated data. Kifer and Machanavajjhala confirmed that correlated data is likely to leak more unexpected privacy \cite{kifer2011no} under differential privacy. In order to break the limitations of correlated differential privacy, they proposed a new privacy model called Pufferfish which can provide different privacy definitions to the needs of customized applications. Pufferfish is quite different from differential privacy, and this privacy model includes the protected target,  background knowledge of the adversary and the neighboring dataset.

\subsection{Pufferfish}
In Pufferfish, three components are used to specify the privacy requirements: $  \mathcal{S}$, a set of secrets that are needed to be protected; $ \mathcal{Q} $, a set of secret pairs that need to be indistinguishable to the adversary; $\Theta $, a class of distributions that represents how the data are generated. The definition of Pufferfish is described as follows.
\begin{Def}
	($ \epsilon $-Pufferfish) \cite{Kifer2014}. A privacy mechanism $ \mathcal{M} $ satisfies $ \epsilon $-Pufferfish in a framework  $(\mathcal{S},\mathcal{Q},{\Theta }) $ if for datasets $ X$ $\sim$ $\theta $ and for all secret pairs$ (s_{i},s_{j}) \in \mathcal{Q}  $ and for all possible output $ \omega \in \Omega $,
\begin{equation}
	 e^{-\epsilon} \cdot \frac{P(s_{i}|\theta)}{p(s_{j}|\theta)} \leq \frac{P(s_{i}|\mathcal{M}(X)=w,\theta)}{P(s_{j}|\mathcal{M}(X)=w,\theta)} \leq e^{-\epsilon} \cdot \frac{P(s_{i}|\theta)}{p(s_{j}|\theta)}  
\end{equation}
	where $\theta$ ($\theta \in \Theta $)  is to represent a probability distribution which denotes the attacker's probabilistic belief and background knowledge. $ P(s_{i}| \theta)  $ and $ P(s_{j}| \theta)  $ are conditional probabilities and the attacker has uncertainly about $ s_{i} $ and $ s_{j} $ ($ P(s_{i}| \theta) \neq0 $, $ P(s_{j}| \theta) \neq0 $). When the $\epsilon$ is small, seeing the sanitized output $ w $ leaks nearly no information to the attacker who is trying to figure out whether $ s_{i} $ or $ s_{j} $ is true.	
\end{Def}

There are two advantages of why Pufferfish privacy framework is able to deal with correlated data: (1) Pufferfish is able to hide private information against data correlation in the dataset since data correlation is assumed to be specified. (2) Pufferfish is capable of dealing with a large number of correlated records, and it can provide a high utility of data. This is because the sensitive information is specified, and various discriminative pairs can be used to protect the sensitive information.

$\mathbf{Pufferfish\ vs\ DP}$	Three main differences are between Pufferfish and differential privacy. (1) The information that we want to protect in Pufferfish is specified and can be various information, while the information we want to protect in differential privacy is whether one user (or record) is in the dataset in the $ \epsilon $-DP. (2) The discriminative pairs can be various in Pufferfish, while the discriminative pair can be regarded as "one record is in the dataset" and "one record is not in the dataset" in $ \epsilon $-DP. (3)  Assumptions are made in data generation in Pufferfish, while data are assumed to be independent in differential privacy. When satisfying some conditions, differential privacy can be regarded as a special case of Pufferfish. Hence, Pufferfish is a kind of generalization of differential privacy, which provides rigorous statistical guarantee to prevent the information leakage.

$ \mathbf{BDP \ vs \ Pufferfish } $	Bayesian differential privacy can be considered as a special case of Pufferfish. When the potential secrets to be the set of all possible values of records in the dataset and discriminative pairs to be the corresponding set of all pairs of secrets, and data are generated by the Bayesian network, and then Pufferfish transforms into Bayesian differential privacy. 
\subsection{Blowfish}
Based on the framework of Pufferfish, another privacy model Blowfish privacy is proposed to provide a rich interface for implementation \cite{he2014blowfish}.
The key feature of Blowfish is a policy that the sensitive information is specified, and adversary knowledge is in the form of a set of deterministic constraints $ Q $ that are known by the public.  With these policies,  mechanisms can be expected to permit more utility since not all properties of an individual need to be kept secret and adversarial attacks on correlated records can be limited due to public known constraints. 
The definition of Blowfish is given below.
\begin{Def}
	($ (\epsilon, P) $ Blowfish) Given a privacy budget $ \epsilon $ and a policy $ P(\mathcal{T,G,I_{Q}}),$ a randomized mechanism $ \mathcal{M} $ satisfies ($ \epsilon,P $)-Blowfish privacy if for any pairs of neighboring datasets $ D $, $ D' $ and for all possible outputs $ \omega \subset \Omega $, we have
	\begin{equation}
	 Pr[\mathcal{M}(D\in \Omega) \leq e^{\epsilon}Pr[\mathcal{M}(D'\in \Omega)] 
	 \end{equation}
	Here, the policy $ P(\mathcal{T,G,I_{Q}})$ is a new notion proposed in Blowfish. For a policy $ P(\mathcal{T,G,I_{Q}}) $, $ \mathcal{T}$ is the domain of the dataset; $ \mathcal{G}=(V,E) $ is a discriminative graph used to present the secret pairs, in which $ V \subset \mathcal{T} $ and $ E \subset \mathcal{T*T} $ denotes values in the domain that an adversary must not distinguish between them; $\mathcal{I_{Q}} $ denoting the set of datasets that are possible under the publicly known constraints $Q$.

\end{Def}


$\mathbf{Blowfish\ vs\ Pufferfish}$  Blowfish borrows the notion of a set of specified secrets that need protection from Pufferfish. In Pufferfish, the adversary knowledge is defined as the assumption about how data are generated, and it tends to be described by the probabilistic correlation function. In Blowfish, the knowledge of correlation is defined as a set of publicly known constraints. This indicates that Blowfish without constraints is equivalent to Pufferfish in the case of adversaries who believe records in the dataset are independent. 
Both Pufferfish and Blowfish are helpful for the data publisher who can customize privacy definitions by carefully defining sensitive information and background knowledge.

$\mathbf{Blowfish\ vs\ DP}$ Differential privacy can be considered as a special case of Blowfish when two conditions are satisfied: (1) The second parameter of the policy $ \mathcal{G} $ is the complete graph on the domain, instead of a part of the domain. (2) There is no publicly known constrains on the dataset. 

\subsection{Mechanisms for Pufferfish}
\subsubsection{The  Wasserstein mechanism }
Every privacy model needs corresponding mechanisms to perform. Some mechanisms proposed are proposed to implement  Pufferfish.
Since there is no general mechanism for the framework of Pufferfish, Wang and Song firstly proposed general mechanisms that can apply in the Pufferfish \cite{wang2016privacy,song2017pufferfish}. They used the Wasserstein distance as a metric to measure the maximum distance between distributions $ P(\mathcal{M}(X)|s_{j},\theta) $ and  $ P(\mathcal{M}(X)|s_{i},\theta) $ for a secret pair $ (s_{i},s_{j}) $. Here, the maximum distance for a secret pair is similar to the maximal difference between the query result on neighboring datasets in differential privacy. Hence, the goal of the Wasserstein mechanism is to measure the maximum distance for a secret pair. The definition of maximum distance is given below.
\begin{Def}
	($\infty$-Wasserstein distance) Suppose for some $ (s_{i},s_{j}) $ and $ \theta $, $ P(f(X)|s_{i},\theta) $ can be transformed into $ P(f(X)|s_{j},\theta) $. Then the maximum distance of two probability mass function is Wassertein distance which is given as
	\begin{equation}
	 W_{i,j,\theta}=W_{\infty}(P(f(x)|s_{i},\theta), P(f(x)|s_{j},\theta))  
	\end{equation}
	where $ f $ is the query.
\end{Def}

Adding Laplace noise with the scale of $ W_{i,j,\theta} $ to the query answers will guarantee the odds ratio of $ s_{i} $ to $ s_{j} $ in the range of $ [e_{-\epsilon},e_{\epsilon}] $. The odds ratio of $ s_{i} $ to $ s_{j} $ is  the probability of $ s_{i} $ being $ s_{j} $ after the attacker seeing the sanitized output. After iterating all pairs $ (s_{i},s_{j})\in \mathcal{Q}$ and all $ \theta \in \Theta $,  the maximum Wassertein distance can be obtained. In Wassertein mechanism, the amount of noise added to the correlated dataset is similar to the form in Laplace mechanism,  
\begin{equation}
 Z=Lap(\frac{W}{\epsilon})  
\end{equation}
where $ W=sup_{(s_i,s_j) \in \mathcal{Q},\theta \in \Theta} W_{\infty}(u_{i,\theta},u_{j,\theta})$.


\subsubsection{The Markov Quilt mechanism}
As Wassertein mechanism may have a complex computation, another mechanism called the Markov Quilt mechanism, based on Bayesian network is proposed for Pufferfish. As we mentioned in Section \RNum{6}, Bayesian network is a popular method to describe data correlation. Hence, there is also a mechanism which uses Bayesian network to design mechanism in Pufferfish.

The mechanism will attempt to find a proper set $ X_A $ such that $ X_{i}$ has low max-influence on $X_A$ under $\Theta$. Here, $ X_A $ can be regarded as a set of nodes that have correlation with $ X_i $. First, we need to quantify the extent of changing the value of a variable $ X_i \in X $ can affect a set of nodes $ X_A \in X $. The maximum influence of the variable $ X_i $ on a set of variables  $ X_A $ is defined as
\begin{equation}
 e(X_A|X_i)=\max \limits_{a,b \in X} \sup \limits_{\theta \in \Theta x_A \in X}log\frac{P(X_A=x_A|X_i=a,\theta)}{P(X_A=x_A|X_i=b,\theta)}
\end{equation}
Hence, the maximum influence is the maximum divergence between distributions $ P((X_A=x_A|X_i=a,\theta)) $ and $ P((X_A=x_A|X_i=b,\theta)) $. In order to find the set $ X_A $ efficiently, Markov Quilt is proposed to find the set and the definition is given below.
\begin{Def}
	(Markov Quilt) A set of nodes $ X_Q $ is Markov Quilt set for a node $ X_i $ if the following conditions are satisfied in the Bayesian network $ G=(X,E) $. (1) Deleting the $ X_G $ can separate $ G $ into two sets $ X_N $ and $ X_R $ and thus $ X=X_N \cup X_R \cup X_Q $ and $ X_i \in X_N $. (2) $X_R$ is independent of $X_i$ conditioned on $X_Q$.
\end{Def}
The main insight behind the Markov Quilt mechanism is that if $ X_i $ and $ X_j $ are distant from each other, then $ X_j $ is largely independent of $ X_i $. Thus, adding noise to the local nodes can obscure the effect of $ X_i $ in the query result. Using Markov Quilt, it is efficient to find $X_R$ which is a set of remote nodes far from $ X_i $ and $ X_N $ which is a set of local nodes near $ X_i $.

\subsection{Discussion of Pufferfish}

The drawback of differential privacy is that it is not enough to erase the participation of a single individual's private value when there are multiple records correlated with each other. Hence, another privacy model, Pufferfish is proposed to cope with the issue of privacy loss on correlated data. 
 The key of Pufferfish is that it considers how the data are generated and the knowledge of potential attackers. Inspired by Pufferfish, more privacy models study the privacy leakage with the consideration of the background knowledge of data generation. Also, increasing mechanisms for Pufferfish are proposed for the application of this privacy model.


\section{Future Directions}
In this section, we will introduce some promising future directions on correlated differential privacy.
In the section \RNum{5}-\RNum{7}, we summarize most works on correlated data under different privacy models. Three lines include: 1)  describing data correlation with correlation parameters, and 2) using correlation models to describe complex data correlation under the setting of differential privacy, and 3) using the framework of Pufferfish to measure data correlation. However, there are still some issues on correlated data that have not been considered yet. 
\subsection{Correlated differential privacy in machine learning}
Differential privacy is privacy model also used in artificial intelligence to prevent data leakage \cite{Ye2019,zhu2019applying,9158374}, especially in machine learning \cite{zhu2018differentially,yang2018machine}.  
Chaudhuri et al. provided an output perturbation \cite{Chaudhuri2009} and objective perturbation mechanism \cite{Chaudhuri2011}. Abadi et al. studied differentially private stochastic gradient descent mechanisms, where noise is added to gradients \cite{Abadi2016}.
However, data correlation has not been considered when adding noise during the learning. Data correlation in the training data is likely to lead to more changes on the training result,  and consequently the adversary is able to obtain more information.
So far, Zhang et al. have proposed a feature selection method to reduce data correlation in the training dataset \cite{zhang2019correlated}. Privacy loss due to data correlation in machine learning still have some open issues, such as quantifying the privacy loss due to data correlation.
\subsection{Multiple correlated relationships}
Real-world datasets are likely to be multiple relationships between records. As the example illustrated in the Section \RNum{4}, two types of relationships can be in the location dataset, such as the user mobility pattern and the social relationship. 
In previous studies, researches assumed that  only one correlation is in the dataset, which is not practical. One intuitive way to cope with multiple correlations in the dataset is to treat different kinds of correlations as the same correlation. And then methods illustrated in Section \RNum{5}-\RNum{7} can be used to deal with the correlated records and protect privacy leakage. Obviously, it is not an optimal solution because the number of correlated records are enlarged and more noise will be added to the dataset. Hence, the utility of datasets will not be desirable. The method to model multiple correlated relationships and the mechanism to guarantee differential privacy under multiple correlations are open issues that need to be explored in the future.

\subsection{Correlations in different datasets}
Currently, most research focuses on the issue of correlated data in a dataset, while correlated data can be distributed in different datasets. The sensitive information may be leaked when multiple entities publish their data sequentially. If the adversary has enough background information of the dataset, the privacy level of these datasets will be degraded especially when records are correlated in different datasets. The privacy level of datasets not only depends on its privacy parameter, but also depends on the privacy parameter of its neighboring datasets. Most studies may be not applicable when correlated data are in different datasets, like the framework of Pufferfish. This is because there are multiple entities, and they need to negotiate with each other and then make the best choices according to each publisher's privacy request and the utility of whole datasets. Wu. et al constructed a game model of multiple players and study the uniqueness of pure Nash Equilibrium \cite{wu2017game}. However, there are still many issues that need to be considered, like the weight of each publisher and each publisher's own privacy requirement. One promising method of this issue can be modeled as a multi-agent systems to achieve the optimal data utility for multiple data entities.

\subsection{Continuous query release}  
When a dataset deals with a large number of queries, data privacy is more vulnerable since the adversary may infer more information via multiple queries. This will leak more information, especially when the records in the dataset are not independent. Zhu et al. studied  the continuous query release for the correlated dataset \cite{Zhu2018}. An iterative-based mechanism \cite{hardt2010multiplicative} is adopted to answer a set of queries on the correlated datasets. During the process of continuous queries, when a query finds an obvious difference between the current dataset and true dataset, the mechanism will have an update on the current dataset in next query. 
Continual query release is a difficult topic in privacy preserving, especially for correlated datasets.  There are still many unsolved problems, e.g., how to deal with various types of statistical queries, and how to incorporate with multiple correlations for continuous query release.

\subsection{Inference attacks on correlated data}
As we mentioned, data correlation in datasets can leak more information than expected when using differential privacy. This makes inference attacks more easier on correlated datasets.
Even though strong protection provided by differential privacy obfuscates the original data using stochastic noise to avoid privacy leakage, privacy leakage is  still breached by some inference attacks.
Shao et al. proposed a novel location inference attack framework, which is able to recover multiple trajectories from differentially private trajectory data using the structured sparsity model \cite{shao2020structured}. In the future, more and more attacks are aiming on correlated differential privacy, and how to defend these attacks is a challenging topic.


\section{Conclusion}
This paper presents a survey on correlated data under different privacy models. Since correlated dataset are expected to leak more privacy than expected, many works focus on how to address with this issue. Basically, these research are mainly classified into three streams: the first focuses on how to use parameters to describe the correlation in differential privacy; the second uses correlation models to describe data correlation in differential privacy; the last method is a new privacy model, called Pufferfish to protect data privacy while keeps a good utility of datasets.  In the first two lines, we analyze different correlation parameters and correlation models of how to describe data correlation, and compare cons and pros of these methods. Simple data correlation can be described by correlation parameters, and complex data correlation can be described by correlation models.
 In the last research line, we analyze Pufferfish, compare the difference of this model with differential privacy, and present mechanisms for Pufferfish. Our goal is to provide an overview of existing work on the issue of correlated dataset. Lastly, we propose some interesting issues that have not been studied or solved in correlated differential privacy. 

\section*{acknowledgements}
This work is supported by an ARC Discovery Project (DP190100981, DP200100946) from the
Australian Research Council, Australia.

\section*{Conflicts of Interest}
The authors declare that there are no conflicts of interest regarding the publication of this paper.

\bibliography{ref}%

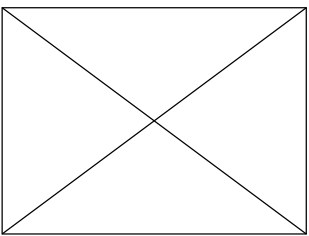
\end{document}